\documentclass{article}

\usepackage{main}
\usepackage[dvipsnames]{xcolor}

\usepackage[utf8]{inputenc} 
\usepackage[T1]{fontenc}    
\usepackage{hyperref}       
\usepackage{url}            
\usepackage{booktabs}       
\usepackage{amsfonts}       
\usepackage{nicefrac}       
\usepackage{microtype}
\usepackage{multirow}
\usepackage{lipsum}
\usepackage{float}
\usepackage{amsmath}
\usepackage{fancyhdr}       
\usepackage{graphicx}       

\title{Physical Momentum in the Indian Stock Market}

\author{
  Tulasi Narendra Das Tripurana\\
  Indian Institute of Technology, Gandhinagar\\
  Gandhinagar\\
  \texttt{tulasi.narendra@btech2015.iitgn.ac.in} \\
   \And
  Naresh Kumar Devulapally \\
  The State University of New York at Buffalo \\
  Buffalo, New York\\
  \texttt{devulapa@buffalo.edu} \\
}

\begin{document}
\maketitle

\begin{abstract}
Our study focuses on determining the presence of abnormal returns for physical momentum portfolios in the context of the Indian market. 
The physical momentum portfolios, comprising stocks from the NSE 500, are constructed for the daily, weekly, monthly, and yearly timescales. In the aforementioned timescales, we empirically evaluate the historical returns and varied risk profiles of these portfolios for the years 2014-2021. It has been observed that the best-performing physical momentum portfolios from each of the four timescales achieved higher returns and better risk measures when compared to the benchmark NIFTY 50 portfolio. We further find that the high-frequency daily time scale exhibits the strongest reversal in the physical momentum effect, wherein the portfolio yielded a 16-fold profit over the initial investment.
\end{abstract}


\section{Introduction}
\paragraph{}
The Efficient Market Hypothesis (EMH) states that a market is said to be ‘Efficient’ if the price of a security always ‘fully reflects’ all available information about that security \cite{fama1970efficient}. EMH also implies that security always trades at its fair value making it impossible to buy undervalued securities or sell overvalued securities. However, there have been multiple studies \cite{schwert2003anomalies,shleifer1997survey} that have challenged EMH by empirically proving the existence of arbitrage, indicating market inefficiency, and such inefficiencies or inadequacies can be exploited to make consistent profits. Although pricing models such as the Capital Asset Pricing Model imply EMH \cite{sharpe1964capital}, alternative theories allowing the existence of market anomalies, such as the adaptive market hypothesis, have been introduced and well documented \cite{lo2004adaptive,lo2005reconciling}. Statistical arbitrage is one such existing market anomaly that plays a key role and is often used by hedge funds and portfolio managers on Wall Street to design profitable strategies \cite{choi2014physical}. Even though their origins are not well explained, statistical arbitrage opportunities such as momentum, mean reversion, and pairs trading strategies have been proven to be profitable in multiple markets \cite{avellaneda2008statistical}.
\paragraph{}
Momentum, a form of statistical arbitrage, is the tendency for asset prices to continue moving in the same direction along which the asset has moved in the past. The central idea of momentum is that stock prices and individuals alike overreact or underreact to information \cite{de1985does}. If that is the case, then profitable trading strategies can be designed based on the stock’s past returns \cite{jegadeesh2011momentum}. For instance, Jegadeesh and Titman \cite{jegadeesh1993returns} proved that portfolios based on buying past winners and selling past losers, yielded abnormal returns. This strategy involves creating a long-short portfolio, wherein the assets are either bought or sold based on their “momentum” during the lookback period. Momentum for a particular asset is measured using a relative strength index, and this index acts as a ranking system to select and compile assets into winner or loser baskets. Long-short portfolios are then constructed by taking a long position on the winner basket and taking a short position on the loser basket. The price momentum effect has been heavily researched in the equities market, but its presence can also be seen in other markets; currency exchange and crypto-currency market are examples where momentum portfolios resulted in excessive market returns \cite{asness2013value,liu2020dynamical}.
\paragraph{}
Price Momentum effect as an arbitrage opportunity has been the center of attention for both fund managers \cite{grinblatt1995momentum,menkhoff2005use} and academics alike. This interest is mainly due to the fact that despite there being no consensus to explain the source of momentum profits, it has remained, as we shall see below, a profitable strategy across global markets. In the US market, Jegadeesh and Titman \cite{jegadeesh1993returns}, showed that over medium-horizons, a period of 6 months to 12 months, firms show trend continuation, i.e past winners continue to outperform past losers. Rouwenhorst \cite{rouwenhorst1998international} also documented similar momentum behavior in European markets, by following Jegadeesh and Titman’s approach to portfolio construction and analyzing its performance in 12 European countries. Kan and Kirikos \cite{kan1996now} found the presence of a short-run momentum effect in the Canadian market, while Chui, Wei, and Titman \cite{chui2000momentum} reported momentum profits in most of the Asian markets except Japan. By contrast, De Bondt and Thaler \cite{de1985does} introduced a contrarian momentum effect, that suggested a reversal of performance in returns for individual firms. The study showed that over long horizons, a period of three to five years, past losers tend to outperform past winners.
\paragraph{}
In the Indian context, Sehgal and Balakrishnan \cite{sehgal2002contrarian} obtained results that, for the years 1989 to 1999, supported a weak reversal in long-horizon returns provided a year is skipped between portfolio formation and holding period. The same study further proved a strong continuation pattern in short-horizon returns also resulting in significantly higher returns than the long-horizon contrarian strategy. Sehgal \cite{sehgal2006rational} show that the momentum returns which are not explained by the CAPM model are explained by the Fama-French three-factor model. They further suggest that rational sources of momentum profits exist. Ansari and Khan \cite{ansari2012momentum} also reported evidence of momentum which, for the years 1995 to 2006, were in conformity with the results observed by Sehgal and Balakrishnan \cite{sehgal2002contrarian}. Ansari and Khan \cite{ansari2012momentum} found that momentum portfolios based 3-month look back and a 3-month holding period resulted in the highest average monthly returns. Garg and Varshney \cite{garg2015momentum} examined the existence of the momentum effect, for the years 2000 to 2013, in four sectors of the Indian economy. The study considered large-cap stocks of Automotive, Banking, Pharmaceutical, and IT sectors and reported the highest presence of momentum effect in the Pharmaceutical sector, proceeded by Automotive and Banking with similar presence, with the least presence in the IT sector. The study further found that portfolios-based long-horizon returns yielded the most profits in all four sectors of the Indian market. Additionally, Mohapatra and Misra \cite{mohapatra2020momentum} examined the effect of momentum, for the years 2005, to 2015, and reported abnormal returns for medium horizon portfolios, which were in accordance with findings reported by Jegadeesh and Titman \cite{jegadeesh1993returns} for the US market. They however found a stark difference, where in the Indian equity market, momentum reversal was not observed for long-run holding periods of over two to five years.
\paragraph{}
Over the years, multiple research papers have been introduced attempting to develop alternative momentum strategies and testing them across the Global markets. Moskowitz et al. \cite{moskowitz2012time} introduced the concept of time series momentum as an alternative to the traditional cross-section momentum that was introduced by Jegadeesh and Titman \cite{jegadeesh1993returns}.  While the traditional cross-section momentum focuses on the relative returns of the assets, time series momentum focuses on the assets' absolute past return. An autoregressive model was used to predict future excess returns of an asset scaled by volatility based on its lagged past returns scaled by volatility. This predicted future excess return was then used to construct the time series portfolio, and the study found that past 12-month excess returns of an asset return positive profits across 58 asset contracts. George and Hwang \cite{george200452} report momentum profits while using a 52-week high price as a ranking criterion. The study finds that the 52-week high strategy has higher predictive power for future returns than traditional past returns. They further suggest that the nearness of a stock’s price to its 52-week high is information that is easily available to investors, and it is common practice to use it as an “anchor” while assessing an asset's value. Rachev et al. \cite{rachev2007momentum} proposed reward-risk ratios as a criterion to select stocks for the momentum portfolio. In addition to the traditional Sharpe Ratio, the study also used alternative reward-risk ratios such as STARR ratio, and R-ratio to better capture the distributional behavior of price data. These alternative reward-risk portfolios provided better risk-adjusted performance than traditional momentum and traditional Sharpe ratio portfolios. The study further stated that investors who consider cumulative return criterion face heavy-tail distributions and hence higher tail risk when compared to investors who follow alternative reward-risk ratios as a selection criterion.
\paragraph{}
Antonacci \cite{antonacci2012momentum} introduced the concept of dual momentum combining both the relative and absolute momentum of an asset to build a momentum portfolio. The study reported higher risk-adjusted annualized returns than traditional cross-section momentum in the US market. The study used a two-stage selection process, where the assets are selected based on their relative strength momentum and if the selected asset further shows an absolute positive momentum with respect to Treasury bills, only when both conditions are met will the asset selected for the dual momentum portfolio. If the asset fails to show absolute momentum, then the Treasury bill return will be used as a proxy investment, as it can act as a safer alternative and ensures the portfolio is diversified as well. Blitz, Huij, and Martens \cite{blitz2011residual} used residual momentum estimated using Fama and French three-factor model to develop an alternative momentum strategy. The study extended the work of Grundy and Martin \cite{grundy2001understanding} which showed that traditional momentum has substantial exposures to the Fama and French factors. The study showed that using residual momentum significantly reduced dynamic factor exposures of the momentum strategy, resulting in a reduction in the volatility of the strategy. It was further found in this study that residual momentum has similar returns when compared to traditional momentum but at half the volatility, i.e., with roughly double the Sharpe ratio.
\paragraph{}
Choi \cite{choi2014physical} measured the strength of an asset using ‘physical’ momentum as an analogy for price momentum, wherein the behavior of the instrument is assumed to be similar to a one-dimensional particle. The study defined mass and velocity for an asset and used these definitions to calculate the physical momentum of that asset. Choi \cite{choi2014physical} then created portfolios based on physical momentum and established the presence of abnormal profits in the US equity market. In our study, we shall use physical momentum as defined by Choi \cite{choi2014physical} and study its existence and the abnormal profits it entails in the Indian equity market scenario.

\section{Introduction to Physical Momentum}
\subsection{One Dimensional Space}
\paragraph{}
Market information and investor behavior are the main driving forces behind price change for an instrument. The instrument price of any asset class is a time-dependent variable whose value is always positive. However, to establish a price space for an instrument that is analogous to physical space, the instrument price needs to be extended to both the positive and negative lines. To extend the price space, one method is to apply log transformation to the instrument price.
\begin{equation}
    x(t)={\log}S(t)
\end{equation}

Where $x(t)$ is the log price and $S(t)$ is the instrument price. Logarithmic price is a common transformation used by a majority of technical analysts in the field of asset management. The advantage of log price is not just limited to extending the price space, but also capturing the price change in terms of percentage rather than in terms of absolute dollars. The distance between prices on the log scale decreases as the price increases. In other words, a {\$}1 increase in price is less significant if the underlying price is much higher as it corresponds to a lesser percentage change. This ensures that higher percentage moves have more significance in the model when compared to lower percentage moves. Log price further ensures that equivalent price changes are represented by the same number of vertical changes on the scale.
\subsection{Velocity}
\paragraph{}
Now that a mathematical representation of instrument price in one-dimensional space has been established, the velocity of the instrument price can be calculated. Choi suggested that the log return $R(t)$ can be a representation of log price velocity.

\begin{equation*}
\begin{split}
    R(t) &={\frac{{\log}S(t)-{\log}S(t-\Delta t)}{\Delta t}} \\
    &={\frac{x(t)-x(t-\Delta t)}{\Delta t}} \\
    &={\frac{\Delta x(t)}{\Delta t}} \\
    &={\frac{dx(t)}{dt}}=v(t)
\end{split}
\end{equation*}
This relationship between log return and velocity is valid under the assumption that $\Delta t\to$0. In our case the assumption holds as the length of our data is large enough that individual time steps can be approximated to be infinitesimally smaller.
\paragraph{}
Cumulative return $r(t)$, can be expressed as,

\begin{equation*}
\begin{split}
    r(t) &={\frac{S(t)-S(t-\Delta t)}{S(t-\Delta t)}} \\
    &={\frac{S(t)}{S(t-\Delta t)}-1} \\
    &=\exp(R(t))-1 \\
    &=\exp(v(t))-1
\end{split}
\end{equation*}
So, the above equation can be re-written as,

\begin{equation}
    v(t)={\log}(r(t)+1)
\end{equation}

\subsection{Mass}
\paragraph{}
The efficient market hypothesis, as discussed previously, implies that all market information of an instrument is available to investors and any change in the price of said instrument is fully explained by any new information available to the investors. Since instrument price is heavily influenced by investor behavior, which in turn, is dictated by market information available to them. In practice, information availability among investors is inhomogeneous, and its effective exchange decides whether the observed price change is meaningful. Every instrument is uniquely affected by its respective investor behavior, raising the need for a metric to capture and normalize the correlation between price and behavior across all assets.

\paragraph{}
In physics, mass is a physical property that is unique to a particle, and given constant force, the mass of a particle decides the particle’s acceleration and velocity. In finance, market information and investor behavior are analogous to physical force, and since log return is the velocity of an instrument price, mass for an instrument can also be established. Choi introduces the concept of financial mass, as an analogy to physical mass, which acts as a filter for market information. When compared to traditional momentum, physical momentum applies mass to amplify the change in log price additionally incorporating instrument-specific market information in the ranking criteria of the momentum strategy.

\paragraph{}
Physical mass and velocity are inversely proportional given constant force or momentum, similarly, the candidate for financial mass should have an inverse relationship with log return. The liquidity of an instrument is a market feature that measures how quickly the instrument can be bought or sold without a significant change in its price. By this definition, it can be inferred that instruments with high liquidity have a greater number of transactions when compared to illiquid instruments. Additionally, liquidity also influences market efficiency, where a highly liquid market is more efficient, resulting in the price change being more meaningful. Datar et al. \cite{datar1998liquidity} showed an inverse relationship between liquidity and future returns, which suggests that liquidity is a possible candidate for financial mass.

\paragraph{}
Liquid markets exhibit tightness, immediacy, breadth, depth, and resiliency \cite{sarr2002measuring}. Volume-based metrics primarily measure the breadth and depth of market liquidity, while bid-ask spreads measure tightness \cite{datar1998liquidity}. Choi considers transaction value, volatility, and volume as measures of liquidity and as financial masses for the physical momentum portfolio. Volume traded is a direct measure of liquidity for an instrument. However, raw volume must be normalized for the ranking criteria in the momentum strategy to be uniform and homogeneous across all assets. This normalization is to account for the fact that assets differ in the amount of total outstanding shares, and hence some assets inherently have higher daily raw volume when compared to other assets. Turnover rate, expressed as $\upsilon$, which is raw volume divided by total outstanding shares, is used as a candidate for financial mass by Choi and in our paper. Additionally, the Inverse Turnover rate, expressed as $1/{\upsilon}$, is used as a candidate for financial mass in our paper.

\paragraph{}
Volatility is the measure of spread for asset returns. Highly volatile stocks have greater variations in stock price compared to low-volatile stocks. Stocks with high volatility grant massive positive returns at the risk of incurring equally massive negative returns. In either case, absolute log returns are positively correlated with volatility. Volatility is an important metric in the field of finance. Since investors are assumed to be risk-averse, changes in the volatility of an instrument can be understood as a direct consequence of efficient information exchange between investors. Since mass is inversely related to log returns, the inverse of volatility, expressed as $1/{\sigma}$, will be used as a financial mass in our paper.

\subsection{Momentum}
With analogies of mass, velocity, and a one-dimensional space for instrument price have been defined, the momentum of the instrument can be calculated. Choi applies three different measures of linear momentum, represented as $ p^1_{t,k}$,$\;p^2_{t,k}$, and $\;p^3_{t,k}$, to calculate the performance of stocks in the US capital markets and rank them to form the physical momentum portfolio.

\begin{equation}
    p_{t,k}^1\left(m,v\right)=\sum_{i=0}^{k-1}{m_{t-i}v_{t-i}}
\end{equation}
\begin{equation}
    p_{t,k}^2\left(m,v\right)=\frac{\sum_{i=0}^{k-1}{m_{t-i}v_{t-i}}}{\sum_{i=0}^{k-1}m_{t-i}}
\end{equation}
\begin{equation}
    p_{t,k}^3\left(m,v\right)=\frac{{\bar{v}}_{t,k}}{\sigma_{t,k}}
\end{equation}
Where $t$, is look back period over which the physical momentum index for the portfolio is calculated, and $k$ is the holding period over which the physical momentum portfolio is held. Mass candidate for the momentum measures $p_{t,k}^1$ and $ \;p_{t,k}^2$, is turnover rate $\upsilon$, while the mass candidate for momentum measure $p_{t,k}^3$ is inverse volatility 1/ $\sigma$.
    
\paragraph{}
$p_{t,k}^3$ is a redefined version of $p_{t,k}^1$, where the mass and velocity for the former are calculated over the lookback period when compared to the latter, where the mass and velocity are calculated at each of the lookback steps. With the above three definitions of momentum, momentum portfolios will be created based on varying lookback period t and holding period k.

\section{Methodology}
\subsection{Portfolio Construction}
\paragraph{}
In our study, we examine the profitability of traditional and contrarian physical momentum strategies for high-frequency, short, medium, and long-horizon time periods. Table~\ref{tab:horizon} maps the different time horizons employed in the study with their respective definitions.

\begin{table}[H]
    \caption{Time steps and their respective horizon category definitions as employed in this study.}
    \centering
    \begin{tabular}{|l|l|l|}
    \hline
    \textbf{Time Steps} & \textbf{Lookback Period} & \textbf{Horizon Category} \\
    \hline
     \textbf{Day}   & 1 to 7 Days    & High Frequency \\
    \hline
     \textbf{Week}  & 2 to 8 Weeks   & Short \\
    \hline
     \textbf{Month} & 3 to 12 Months & Medium \\
    \hline
     \textbf{Year}  & >=1 Years   & Long \\
    \hline
    
    \end{tabular}
    \label{tab:horizon}
\end{table}

Our momentum portfolios are built using Jegadeesh and Titman's $J$-month/$K$-month method, where $J$ is the lookback period in months and $K$ is the holding duration in months. Our method broadens the scope of this strategy by extending it over days, weeks, and years. The portfolio is built at time $t=0$, and the physical momentum, as defined in the preceding section, is determined for all of the stocks in our universe from time $t=-J$ to time $t=-1$. The stocks are then ranked and divided into equal groups based on their momentum values. For example, if the universe contains $500$ stocks, $50$ groups with ten stocks each are constructed. These groups are named using Jegadeesh and Titman’s nomenclature, where the top-ranked group is the winner group and is named $R50$, while the bottom-ranked group is the loser group and is named $R1$. In the traditional strategy, when the portfolio is constructed at $t=0$, the winner group is bought, and the loser group is shorted in equal cash-size positions to create a dollar-neutral portfolio. In the contrarian strategy, the dollar-neutral portfolio at time $t=0$ is constructed by short-selling the winners and buying the losers. In either of these strategies, the portfolio is held until $t=k$ and is liquidated at the end of the holding period by closing the winner and loser positions.

\subsection{High Frequency Portfolios}
\paragraph{}
High-frequency portfolios are constructed based on daily time horizons. The lookback period J, and the holding period K, vary from 1 to 7 days. For each mass candidate in the $p_{t,k}^1$, $\;p_{t,k}^2$ momentum criteria, a total of 49 portfolios are created. For $p_{t,k}^3$ momentum, however, this is not the case because the lookback period J can only take values from 2 to 7 days. This is because the mass of $p_{t,k}^3$ momentum is defined by historical standard deviation, and a singular lookback period has no standard deviation measure. Based on this limitation, 237 physical momentum portfolios with a time horizon of days are constructed. These high-frequency portfolios are built every day at market open and liquidated at market close on the last day of the holding period, assuming both days are trading days.

\subsection{Short Horizon Portfolios}
\paragraph{}
Short-horizon portfolios are constructed based on weekly time horizons. The lookback period J varies from 2 to 8 weeks, while the holding period K, varies from 1 to 8 weeks. For each mass candidate in the $p_{t,k}^1$, $\;p_{t,k}^2$, and $\;p_{t,k}^3$ momentum definitions, 56 portfolios can be constructed. This generates a total of 280 portfolios with weeks as their time horizon. These short-horizon portfolios are constructed on Monday or the first valid trading day of every week and liquidated on the last valid trading day of their respective holding period.

\subsection{Medium Horizon Portfolios}
\paragraph{}
Medium horizon portfolios are constructed based on monthly time horizons. The lookback period J varies from 3 to 12 months, while the holding period K, varies from 1 to 12 months. For each mass candidate in the $p_{t,k}^1$, $\;p_{t,k}^2$, and $\;p_{t,k}^3$ momentum definitions, a total of 120 portfolios can be constructed. This generates a total of 600 portfolios with months as their time horizon. The first valid trading day of each month is used to build these medium horizon portfolios, which are then liquidated at the conclusion of their respective holding periods.

\subsection{Long Horizon Portfolios}
\paragraph{}
Long-horizon portfolios are constructed based on yearly time horizons. The lookback period J varies from 2 to 5 years, while the holding period K, varies from 1 to 3 years. Since the length of our time period is limited to eight years, combinations of J/K whose sum exceeds eight years will be excluded. Based on this limitation, for each mass candidate in the $p_{t,k}^1$, $\;p_{t,k}^2$, and $\;p_{t,k}^3$ momentum definitions, 12 portfolios can be constructed. This generates a total of 60 portfolios with years as their time horizon. These long-horizon portfolios are constructed on the first valid trading day of every year and liquidated at the end of their respective holding period.

\subsection{Overlapping Portfolios}
\paragraph{}
Overlapping portfolios are a major aspect in the Jegadeesh and Titman \cite{jegadeesh2011momentum} study of momentum. For multi-period holding strategies, i.e., strategies whose $K>1$, there exist portfolios that overlap over a period of time. This is due to the fact that portfolios created in previous $K-1$ steps have not yet been liquidated when the new portfolio is created. This indicates that at any given time t, between $t=0$ and $t=K$, the number of portfolios that are held simultaneously is equal to the holding period K. {The following diagram illustrates the number of portfolios that are held at time t, for $J=2$ months and $K=3$ months strategy.}

According to Jegadeesh and Titman \cite{jegadeesh2011momentum}, there is no significant difference in returns between overlapping and non-overlapping portfolios. In addition to this, overlapping portfolios provide the added benefit of diversification. Since our dataset is limited to eight years, constructing non-overlapping portfolios would yield fewer return samples when compared to overlapping portfolios.

\subsection{Portfolio Performance Measurement}
\paragraph{}
To calculate the performance of a zero-cost portfolio, both the long and short portfolios are assumed to be bought, and their respective returns are calculated. The return of the short portfolio is then subtracted from the return of the long portfolio to calculate the zero-cost portfolio return. The winners are bought in a traditional momentum strategy, while the losers are shorted. Hence, the return of a traditional momentum portfolio, represented as $R_p$ is calculated as,
\begin{equation*}
    R_p=R_w - R_l
\end{equation*}
where $R_w$ and $R_l$ are the expected returns of the winner and loser groups, respectively. For the contrarian momentum strategy, the winners are shorted, and the losers are bought. Hence, the return of a contrarian momentum portfolio, represented as $R^*_p$ is,

\begin{equation*}
    R^*_p=R_l-R_w
\end{equation*}
In our study, the portfolios constructed are paper traded and are not implemented in the real market, hence commissions are not considered while calculating momentum profits.

\subsection{Data Tabulation and Benchmark}
\paragraph{}
A total of 1177 traditional and contrarian momentum portfolios are constructed based on the above-defined time horizons. The results section is divided into four subsections, to discuss the findings for all the above-shown time horizons. For every momentum definition in each of the time horizons, the best-performing portfolio’s statistics and risk measures are tabulated and benchmarked against Nifty 50. Nifty 50 is a well-diversified index consisting of 50 companies and is the most common benchmark used by asset managers in India. Additionally, the familiar CAPM equation will be used to estimate and provide support for the extranormal returns from each of the best-performing physical momentum portfolios.

\section{Results}
\subsection{High Frequency Returns}
\paragraph{}
Table~\ref{tab:daily_returns} contains all high-frequency monthly mean returns and standard deviations of the best-performing portfolios from every mass and momentum combination. All the high-frequency physical momentum portfolios have a higher mean return than the benchmark Nifty index, with $p{\ast}^2(1/\upsilon,R)$ portfolio having the highest monthly mean return of 6.04\%. Additionally, the $p{\ast}^2(1/\upsilon,R)$ portfolio has lower volatility (4.78\%) than the Nifty benchmark (5.03\%).

\begin{table}[H]
    \caption{Returns and standard deviation measures of high-frequency physical momentum portfolios in NSE 500.}
    \centering
    \begin{tabular}{|l|l|l|l|l|l|}
    
    \hline
    Portfolio & Strategy & J-K & Basket & Mean & Std. Dev. \\
    \hline
    
    \multirow{3}{*}{$p{\ast}^1(\upsilon,R)$} &	\multirow{3}{*}{Contrarian} & \multirow{3}{*}{1-2} & Winner (W) & -4.64415 & 7.178647 \\ \cline{4-6} 
                                             &                              &                      & Loser (L)  & -1.65496 & 7.387066 \\ \cline{4-6}
                                             &                              &                      & L - W      & 2.989188 & 2.61962 \\
    \hline

    \multirow{3}{*}{$p^1(1/\upsilon,R)$} &	\multirow{3}{*}{Traditional} & \multirow{3}{*}{1-2} & Winner (W) & -0.90117 & 6.522656 \\ \cline{4-6} 
                                         &                               &                      & Loser (L)  & -2.34479 & 5.579123 \\ \cline{4-6}
                                         &                               &                      & W - L      & 1.443627 & 2.420286 \\
    \hline

    \multirow{3}{*}{$p{\ast}^2(\upsilon,R)$} &	\multirow{3}{*}{Contrarian} & \multirow{3}{*}{2-1}  & Winner (W) & -9.46681  & 7.500539  \\ \cline{4-6} 
                                             &                               &                      & Loser (L)  & -4.93755  & 7.915264 \\ \cline{4-6}
                                             &                               &                      & L - W      &  4.529257 & 4.696118 \\
    \hline

    \multirow{3}{*}{$p{\ast}^2(1/\upsilon,R)$} & \multirow{3}{*}{Contrarian} & \multirow{3}{*}{3-1}  & Winner (W) & -8.98657  & 7.605155  \\ \cline{4-6} 
                                             &                               &                      & Loser (L)  & -2.94504  & 7.076357 \\ \cline{4-6}
                                             &                               &                      & L - W      &  6.041535 & 4.780503 \\
    \hline

    \multirow{3}{*}{$p{\ast}^3(1/\sigma,R)$} & \multirow{3}{*}{Contrarian} & \multirow{3}{*}{3-1}  & Winner (W) & -4.51109  & 5.747774  \\ \cline{4-6} 
                                             &                               &                      & Loser (L)  & -1.65428  & 6.08766 \\ \cline{4-6}
                                             &                               &                      & L - W      &  2.856803 & 3.646641 \\
    \hline
    \end{tabular}
    \label{tab:daily_returns}
\end{table}

\paragraph{}
All of the best-performing portfolios, with the exception of $p^1\left(1/\upsilon,R\right)$,  are contrarian establishing a reversal in high-frequency returns.$p^1(1/\upsilon,R)$ momentum is the only non-contrarian portfolio whose monthly mean return beat the benchmark. However, $p^1(1/\upsilon,R)$ also is the only portfolio with the least monthly mean returns and standard deviation. The winner and loser groups of all five physical momentum portfolios have negative mean returns, regardless of the strategy. This suggests that the loser group in all momentum criteria shows stronger momentum continuation, while the winner groups show a strong reversal. This additionally proves that the profits gained by dollar-neutral portfolios based on the traditional (contrarian) approach are purely attributable to the winner (loser) group relatively outperforming the loser (winner) group.

\paragraph{}
It can also be seen that dollar-neutral portfolios exhibit lower volatilities when compared to the individual winner and loser groups that constitute them. This reduction in volatility can be attributed to the diversification effect introduced by the assets of the winner and loser groups. P2 portfolios outperformed both P1 and P3 portfolios in terms of mean returns. This is because the winner groups of P2 contrarian portfolios show stronger reversals. However, P2 portfolios exhibit twice the volatility when compared to P1 and P3 portfolios making them riskier. The P3 portfolio has the least return among contrarian portfolios and relatively higher volatility, making it unsuitable for investment.

\begin{table}[H]
    \caption{Risk measures of high-frequency physical momentum portfolios in NSE 500.}
    \centering
    \begin{tabular}{|l|l|l|l|l|l|l|l|}
    
    \hline
    Portfolio & Strategy & J-K & Basket & Fin. Wealth & Sharpe Ratio & VaR$_{95\%}$ & MDD\\
    \hline
    
    $p{\ast}^1(\upsilon,R)$ &	Contrarian   & 1-2 & (L - W) & 4.117029 &  0.550245 &  3.028288 & -12.5527 \\ 
    \hline

    $p^1(1/\upsilon,R)$     &	Traditional  & 1-2 & (W - L) & 1.906175 &  0.295709	&  2.57145	&  -14.8726 \\ 
                
    \hline

    $p{\ast}^2(\upsilon,R)$ &	Contrarian   & 2-1 & (L - W) & 8.290232	&  0.461488	&  4.423333	& -22.1972 \\
    \hline

    $p{\ast}^2(1/\upsilon,R)$ & Contrarian & 3-1  & (L - W)  & 17.08302	& 0.617544	& 3.513963	& -13.6907 \\
    \hline

    $p{\ast}^3(1/\sigma,R)$ & Contrarian & 3-1  & (L - W)    &  3.689321	& 0.379053	& 4.471247	& -23.4431 \\

    \hline
    \end{tabular}
    \label{tab:daily_risks}
\end{table}

\paragraph{}
Table~\ref{tab:daily_risks} contains the risk measures of all the high-frequency dollar-neutral portfolios. All contrarian strategies have outperformed the benchmark Nifty in every risk metric. The momentum portfolios’ 95\% VaR values range from 2.5\% to 4.7\% and are at least 25\% lower than the benchmark Nifty’s 95\% VaR value. Similarly, the maximum drawdowns range from -12.5\% to -23.4\% and are reduced by a minimum of 20\% when compared to the benchmark Nifty’s -29.3\% drawdown. The outperformance of every contrarian portfolio in terms of final wealth can be explained as a result of adopting lower risk. The exception is $p^1\left(1/\upsilon,R\right)$ portfolio which failed to outperform the benchmark in terms of final wealth even with lower risk measures.

\begin{figure}[ht]
\centering 
{\includegraphics{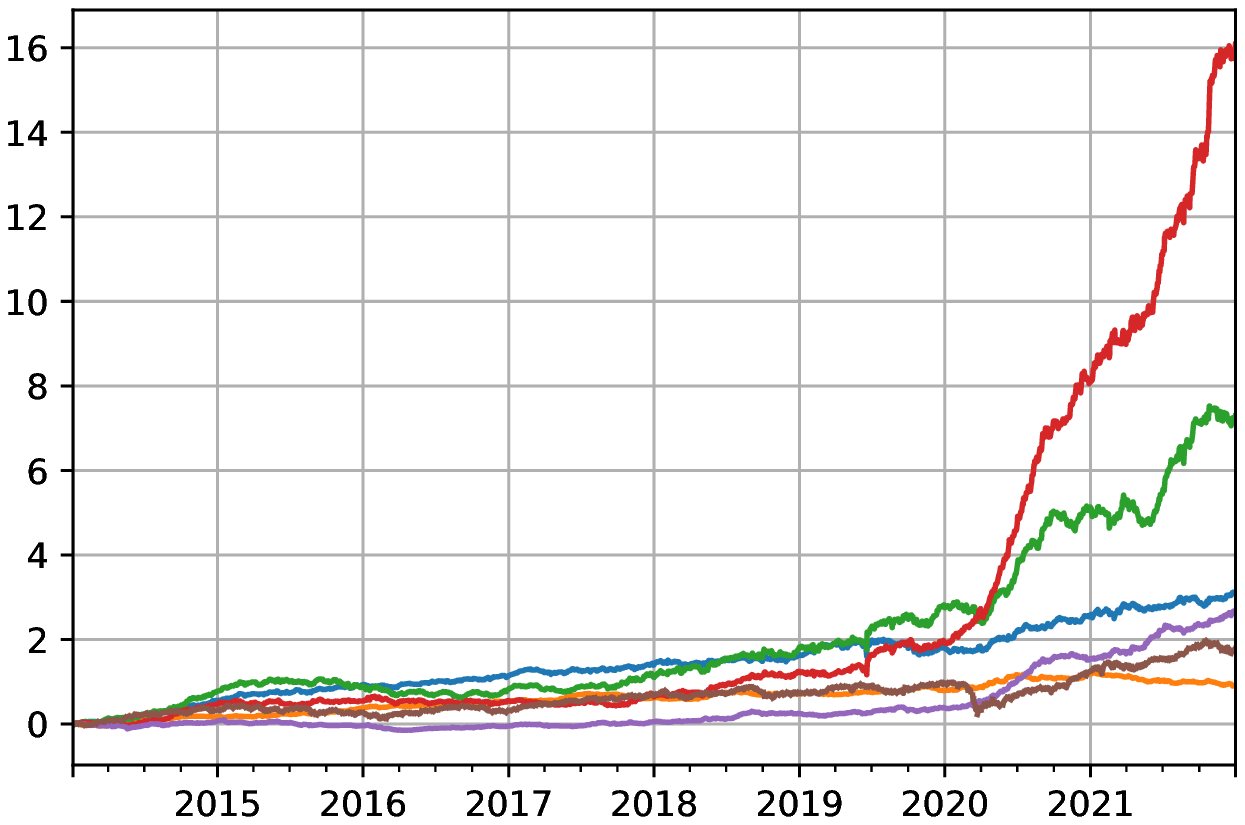}}
\caption{Cumulative returns for $p{\ast}^1\left(\upsilon,R\right)$ (Blue), $p^1\left(1/\upsilon,R\right)$ (Yellow), $p{\ast}^2\left(\upsilon,R\right)$ (Green), $p{\ast}^2\left(1/\upsilon,R\right)$ (Red), $p{\ast}^3\left(1/\sigma,R\right)$ (Purple) high-frequency physical momentum portfolios vs Nifty 50 (Brown) in Indian Market.}
\label{fig:daily}
\end{figure}

\paragraph{}
Figure~\ref{fig:daily} shows the historical daily portfolio values of all high-frequency momentum portfolios along with the Nifty benchmark. Similar to our findings in monthly mean returns, P2 momentum portfolios outperform P1 and P3 portfolios in terms of final wealth. ${p\ast}^2(1/\upsilon,R)$ is the best-performing portfolio yielding a 16-fold increment over the initial investment. From the figure, it can be seen that prior to the Covid-19 market crash in March 2020, the P3 portfolio underperformed when compared to the Nifty benchmark. However, the P3 portfolio managed to outperform the Nifty benchmark by taking advantage of the post-Covid-19 bull run observed in the Indian market \cite{dhall2020covid}. With the exception of $p^1\left(1/\upsilon,R\right)$ portfolio, all the high-frequency momentum portfolios managed to leverage this bull run.

\subsection{Short Horizon Returns}
\paragraph{}
Table~\ref{tab:weekly_returns} contains all short horizon monthly mean returns and standard deviations of the best-performing portfolios from every mass and momentum combination. All the short horizon physical momentum portfolios have a higher mean return than the benchmark Nifty index, with $p^2(1/\upsilon,R)$ portfolio having the highest monthly mean return of 2.54\%. Additionally, the $p^2(1/\upsilon,R)$ portfolio has lower volatility (4.42\%) than the Nifty benchmark (5.03\%).

\begin{table}[]
    \caption{Returns and standard deviation measures of short horizon physical momentum portfolios in NSE 500.}
    \centering
    \begin{tabular}{|l|l|l|l|l|l|}
    
    \hline
    Portfolio & Strategy & J-K & Basket & Mean & Std. Dev. \\
    \hline
    
    \multirow{3}{*}{$p^1(\upsilon,R)$} &	\multirow{3}{*}{Traditional} & \multirow{3}{*}{7-1} & Winner (W) & 0.174126 & 8.661214 \\ \cline{4-6} 
                                             &                              &                      & Loser (L)  & -1.99152 & 9.896505 \\ \cline{4-6}
                                             &                              &                      & W - L      & 2.16565 & 4.473935 \\
    \hline

    \multirow{3}{*}{$p^1(1/\upsilon,R)$} &	\multirow{3}{*}{Traditional} & \multirow{3}{*}{7-1} & Winner (W) & 0.804623	& 6.30491  \\ \cline{4-6} 
                                         &                               &                      & Loser (L)  & -1.62198 & 6.974194 \\ \cline{4-6}
                                         &                               &                      & W - L      & 2.426601 & 2.895703 \\
    \hline

    \multirow{3}{*}{$p^2(\upsilon,R)$} &	\multirow{3}{*}{Traditional} & \multirow{3}{*}{7-1}  & Winner (W) & -0.72846 &	8.514958  \\ \cline{4-6} 
                                             &                               &                      & Loser (L)  & -2.29441 & 9.877659 \\ \cline{4-6}
                                             &                               &                      & W - L      &  1.565944 & 4.77213 \\
    \hline

    \multirow{3}{*}{$p^2(1/\upsilon,R)$} & \multirow{3}{*}{Traditional} & \multirow{3}{*}{8-8}  & Winner (W) & 1.323656 & 8.422598  \\ \cline{4-6}
                                             &                               &                      & Loser (L)  & -1.22108 &	12.4391 \\ \cline{4-6}
                                             &                               &                      & W - L      &  2.544735 & 4.427298 \\
    \hline

    \multirow{3}{*}{$p^3(1/\sigma,R)$} & \multirow{3}{*}{Traditional} & \multirow{3}{*}{8-8}  & Winner (W) & 1.727272 & 6.182747  \\ \cline{4-6} 
                                             &                               &                      & Loser (L)  & 0.390887 & 8.261543 \\ \cline{4-6}
                                             &                               &                      & W - L      &  1.336385 & 2.838392 \\
    \hline
    \end{tabular}
    \label{tab:weekly_returns}
\end{table}

\paragraph{}
All of the above short-horizon portfolios employ the traditional physical momentum strategy, proving the existence of continuation in short-horizon returns. Additionally, dollar-neutral portfolios show evidence of diversification as they exhibit lower volatilities when compared to the winner and loser groups that constitute them. The P3 portfolio has the least return among the short-horizon portfolios. This is due to the presence of a weak reversal in returns of the P3 loser group, while the P1 and P2 loser groups showed strong continuation. It is also clear that the P1 and P2 portfolios' gains were primarily attributable to the loser group's significant continuation of returns, whilst their respective winner groups demonstrated weak continuation or reversal.

\begin{table}[]
    \caption{Risk measures of short horizon physical momentum portfolios in NSE 500.}
    \centering
    \begin{tabular}{|l|l|l|l|l|l|l|l|}
    
    \hline
    Portfolio & Strategy & J-K & Basket & Fin. Wealth & Sharpe Ratio & VaR$_{95\%}$ & MDD\\
    \hline
    
    $p^1(\upsilon,R)$ &	Traditional   & 7-1 & (W - L) & 2.506523 & 0.213385	& 6.107373 & -20.3576 \\ 
    \hline

    $p^1(1/\upsilon,R)$ & Traditional  & 7-1 & (W - L) & 2.989653 & 0.394423 & 3.486973 & -9.43847 \\ 
                
    \hline

    $p^2(\upsilon,R)$ &	Traditional   & 7-1 & (W - L) & 1.857258 & 0.136727 & 7.051863 & -25.9175 \\
    \hline

    $p^2(1/\upsilon,R)$ & Traditional & 8-8  & (W - L)  & 2.691209 & 0.236764 & 4.677592 & -18.4447 \\
    \hline

    $p^3(1/\sigma,R)$ & Traditional & 8-8  & (W - L)   &  1.699671 & 0.203426 &	3.423759 & -11.0897 \\

    \hline
    \end{tabular}
    \label{tab:weekly_risks}
\end{table}

\begin{figure}[ht]
\centering 
{\includegraphics{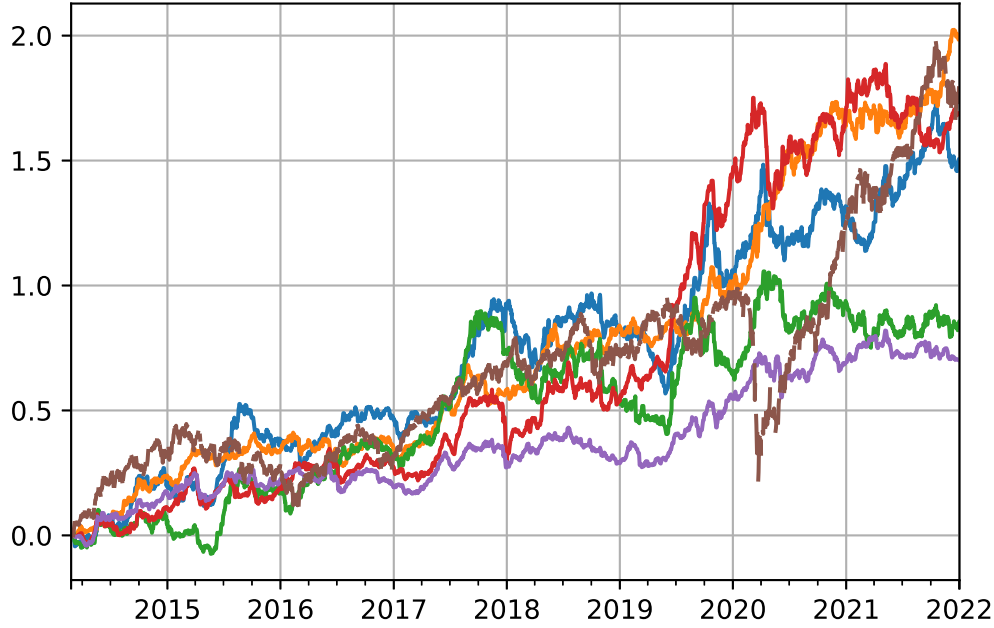}}
\caption{Cumulative returns for $p^1\left(\upsilon,R\right)$ (Blue), $p^1\left(1/\upsilon,R\right)$ (Yellow), $p^2\left(\upsilon,R\right)$ (Green), $p^2\left(1/\upsilon,R\right)$ (Red), $p^3\left(1/\sigma,R\right)$ (Purple) short horizon physical momentum portfolios vs Nifty 50 (Brown) in Indian Market.}
\label{fig:weekly}
\end{figure}

\paragraph{}
Table~\ref{tab:weekly_risks} contains the risk measures of all the short-horizon dollar-neutral momentum portfolios. $p^1(1/\upsilon,R)$ portfolio is the only momentum portfolio that managed to beat the benchmark nifty index in terms of risk measures. Even though $p^2(1/\upsilon,R)$ portfolio’s mean returns (2.54\%) were higher than $p^1(1/\upsilon,R)$ portfolio’s mean returns (2.42\%), the $p^2(1/\upsilon,R)$ portfolio failed to outperform both Nifty and $p^1(1/\upsilon,R)$ in terms of final wealth. This can be attributed to the $p^2(1/\upsilon,R)$ portfolio’s higher volatility, causing an increase in risk measures, primarily a two-fold increase in maximum drawdown when compared to the  $p^1(1/\upsilon,R)$ portfolio. The maximum drawdowns range from -11.08\% to -25.91\% and are reduced by a minimum of 11.6\% when compared to the benchmark Nifty’s -29.34\% drawdown.
\paragraph{}
Figure~\ref{fig:weekly} shows the historical daily portfolio values of all short-horizon momentum portfolios along with the Nifty benchmark. Similar to our findings in monthly mean returns, the P3 momentum portfolio was the worst-performing portfolio in terms of final wealth. Whereas $p^1(1/\upsilon,R)$ is the best-performing portfolio yielding nearly a three-fold increment over the initial investment. From the figure above, it can be seen for P1 and P2 portfolios that employed Inverse Turnover Rate as their mass, namely $p^1(1/\upsilon,R)$ and $p^2(1/\upsilon,R)$, reacted the most to the Covid-19 market crash when compared to the remaining portfolios. With the exception of $p^1(1/\upsilon,R)$, none of the short horizon momentum portfolios managed to leverage the post-Covid-19 Indian market bull run \cite{dhall2020covid}.

\subsection{Medium horizon Returns}
\paragraph{}
For medium horizon momentum portfolios, we shall benchmark the results against Nifty for the period 2015-2021. This is because the lookback period for the best-performing medium horizon portfolios ranges between 6 to 12 months. This shifts the investing start date for every medium horizon portfolio, and the benchmark statistics and risk values will shift accordingly. Since one of the best-performing portfolios has a maximum lookback duration of 12 months, subtracting one year from our test period should suffice, and so 2015-2021 is selected as the new test period.

\begin{table}[]
    \caption{Returns and standard deviation measures of medium horizon physical momentum portfolios in NSE 500.}
    \centering
    \begin{tabular}{|l|l|l|l|l|l|}
    
    \hline
    Portfolio & Strategy & J-K & Basket & Mean & Std. Dev. \\
    \hline
    
    \multirow{3}{*}{$p^1(\upsilon,R)$} &	\multirow{3}{*}{Traditional} & \multirow{3}{*}{8-1} & Winner (W) & 1.306097	& 8.180783 \\ \cline{4-6} 
                                             &                              &                      & Loser (L)  & -1.83147 & 13.73708 \\ \cline{4-6}
                                             &                              &                      & W - L      & 3.137563 & 5.075968 \\
    \hline

    \multirow{3}{*}{$p^1(1/\upsilon,R)$} &	\multirow{3}{*}{Traditional} & \multirow{3}{*}{6-2} & Winner (W) & 2.12104 & 7.102213  \\ \cline{4-6}
                                         &                               &                      & Loser (L)  & -0.20381 & 8.403248 \\ \cline{4-6}
                                         &                               &                      & W - L      & 2.32485 & 3.097818 \\
    \hline

    \multirow{3}{*}{$p^2(\upsilon,R)$} &	\multirow{3}{*}{Traditional} & \multirow{3}{*}{9-4}  & Winner (W) & 1.058269 & 8.030754  \\ \cline{4-6}
                                             &                               &                      & Loser (L)  & -1.8857 & 12.8104 \\ \cline{4-6}
                                             &                               &                      & W - L      &  2.943969 & 4.089426 \\
    \hline

    \multirow{3}{*}{$p^2(1/\upsilon,R)$} & \multirow{3}{*}{Traditional} & \multirow{3}{*}{6-2}  & Winner (W) & 2.519064	& 8.551126  \\ \cline{4-6}
                                             &                               &                      & Loser (L)  & -1.80115 & 14.15218 \\ \cline{4-6}
                                             &                               &                      & W - L      &  4.320216	& 5.311912 \\
    \hline

    \multirow{3}{*}{$p^3(1/\sigma,R)$} & \multirow{3}{*}{Traditional} & \multirow{3}{*}{12-4}  & Winner (W) & 1.653944 & 6.185016  \\ \cline{4-6} 
                                             &                               &                      & Loser (L)  & -0.84587 & 11.28948 \\ \cline{4-6}
                                             &                               &                      & W - L      &  2.499812 & 4.466255 \\
    \hline
    \end{tabular}
    \label{tab:monthly_returns}
\end{table}

\paragraph{}
Table~\ref{tab:monthly_returns} contains all medium horizon monthly mean returns and standard deviations of the best-performing portfolios from every mass and momentum combination. All the medium horizon physical momentum portfolios have a higher mean return than the benchmark Nifty index, with $p^2(1/\upsilon,R)$ portfolio having the highest monthly mean return of 4.32\%. However, $p^2(1/\upsilon,R)$ has the highest volatility of 5.31\% and was the only medium horizon portfolio that failed to beat the benchmark volatility. For the $p^2(1/\upsilon,R)$ portfolio, it can be inferred that its high mean return, and the subsequent outperformance can be attributed to its high-risk profile, i.e., its high volatility.

\paragraph{}
All of the above medium horizon portfolios employ the traditional physical momentum strategy, proving the existence of continuation for medium horizon returns. The volatility levels of every loser group are higher than their respective winning group. Additionally, dollar-neutral portfolios show evidence of diversification as they exhibit lower volatilities when compared to the winner and loser groups that constitute them. P2 portfolios are slightly better performing than P1 and P3 portfolios, mainly due to the presence of stronger continuation of returns in P2’s loser groups when compared to P1 and P3 loser groups. Loser groups of P1 and P2 portfolios with Turnover Rate ($\upsilon$) as their mass candidate show similar strength in momentum continuation and outperformed their respective winner groups. Winner groups of P1 and P2 portfolios with Inverse Turnover Rate (1/$\upsilon$) and P3 portfolio with Inverse volatility (1/$\sigma$) significantly outperformed their respective loser groups and contributed the most to their respective dollar-neutral portfolios.

\begin{table}[]
    \caption{Risk measures of medium horizon physical momentum portfolios in NSE 500.}
    \centering
    \begin{tabular}{|l|l|l|l|l|l|l|l|}
    
    \hline
    Portfolio & Strategy & J-K & Basket & Fin. Wealth & Sharpe Ratio & VaR$_{95\%}$ & MDD\\
    \hline
    
    $p^1(\upsilon,R)$ &	Traditional   & 8-1 & (W - L) & 2.674691 & 0.224043 & 9.082746 & -32.4652 \\ 
    \hline

    $p^1(1/\upsilon,R)$ & Traditional  & 6-2 & (W - L) & 2.640953 & 0.361444 & 3.485223 & -10.7785 \\ 
                
    \hline

    $p^2(\upsilon,R)$ &	Traditional   & 9-4 & (W - L) & 2.75605 & 0.294026 & 5.568717 & -16.7409 \\
    \hline

    $p^2(1/\upsilon,R)$ & Traditional & 6-2  & (W - L)  & 4.354233 & 0.324809 & 6.799119 & -16.9386 \\
    \hline

    $p^3(1/\sigma,R)$ & Traditional & 12-4  & (W - L)   &  2.36502 & 0.23382 & 4.062821 & -28.0848 \\

    \hline
    \end{tabular}
    \label{tab:monthly_risks}
\end{table}

\paragraph{}
Table~\ref{tab:monthly_risks} contains the risk measures of all the medium horizon dollar-neutral momentum portfolios. All the medium horizon portfolios managed to outperform the benchmark Nifty in terms of final wealth and Sharpe ratio. $P^1(\upsilon,R)$, $p^2(1/\upsilon,R)$ are the only portfolios that failed to outperform the benchmark in terms of 95\% VaR measure, furthermore, $p^1(\upsilon,R)$ failed to outperform in terms of maximum drawdown as well. For P2 portfolios, maximum drawdowns were reduced by nearly 45\% with respect to the benchmark Nifty.

\begin{figure}[ht]
\centering 
{\includegraphics{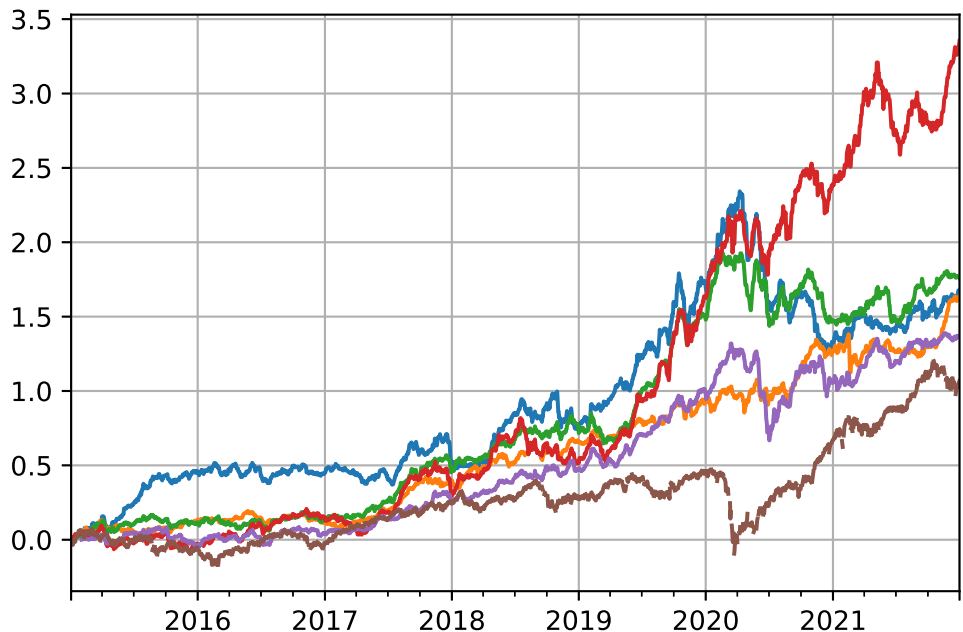}}
\caption{Cumulative returns for $p^1\left(\upsilon,R\right)$ (Blue), $p^1\left(1/\upsilon,R\right)$ (Yellow), $p^2\left(\upsilon,R\right)$ (Green), $p^2\left(1/\upsilon,R\right)$ (Red), $p^3\left(1/\sigma,R\right)$ (Purple) medium horizon physical momentum portfolios vs Nifty 50 (Brown) in Indian Market.}
\label{fig:monthly}
\end{figure}

\paragraph{}
Figure~\ref{fig:monthly} shows the historical daily portfolio values of all medium horizon momentum portfolios along with the Nifty benchmark. The P3 momentum portfolio was the worst-performing portfolio in terms of final wealth, returning 2.3 times the initial investment. Whereas $p^1(1/\upsilon,R)$ is the best-performing portfolio yielding a four-fold return on the initial investment. From the figure above, it can be seen that prior to the Covid-19 market crash in March 2020, portfolios $p^1(1/\upsilon,R)$, $p^3(1/\sigma,R)$ did outperform the benchmark, but the major outperformance in terms of portfolio value occurred during the post-Covid-19 bull run observed in the Indian market \cite{dhall2020covid}. The $p^1(1/\upsilon,R)$ portfolio was the only portfolio that was unaffected by the 2020 market crash, while $p^1(\upsilon,R)$ portfolio was the most affected. Additionally, $p^1(\upsilon,R)$ was the only portfolio that failed to leverage the post-covid-19 bull run.

\subsection{Long Horizon Returns}
\paragraph{}
For long-horizon momentum portfolios, we shall benchmark the results against Nifty for the period 2018-2021. This is because the lookback period for the best-performing long-horizon portfolios ranges between 1 to 3 years. This shifts the investing start date for every medium horizon portfolio, and the benchmark statistics and risk values will shift accordingly. Since one of the best-performing portfolios has a maximum lookback duration of three years, subtracting three years from our test period should suffice, and so 2018-2021 is selected as the new test period.

\begin{table}[]
    \caption{Returns and standard deviation measures of long horizon physical momentum portfolios in NSE 500.}
    \centering
    \begin{tabular}{|l|l|l|l|l|l|}
    
    \hline
    Portfolio & Strategy & J-K & Basket & Mean & Std. Dev. \\
    \hline
    
    \multirow{3}{*}{$p^1(\upsilon,R)$} &	\multirow{3}{*}{Traditional} & \multirow{3}{*}{1-1} & Winner (W) & 0.991022 & 9.229006 \\ \cline{4-6} 
                                             &                              &                      & Loser (L)  & -2.01568 & 16.88496 \\ \cline{4-6}
                                             &                              &                      & W - L      & 3.006704 & 4.197911 \\
    \hline

    \multirow{3}{*}{$p^1(1/\upsilon,R)$} &	\multirow{3}{*}{Traditional} & \multirow{3}{*}{2-1} & Winner (W) & 2.170307 & 6.976874  \\ \cline{4-6} 
                                         &                               &                      & Loser (L)  & -0.92625 & 10.42304 \\ \cline{4-6}
                                         &                               &                      & W - L      & 3.096553 & 4.640399 \\
    \hline

    \multirow{3}{*}{$p^2(\upsilon,R)$} &	\multirow{3}{*}{Traditional} & \multirow{3}{*}{1-1}  & Winner (W) & 1.150526 & 10.50427  \\ \cline{4-6} 
                                             &                               &                      & Loser (L)  & -2.06852 & 16.00266 \\ \cline{4-6}
                                             &                               &                      & W - L      &  3.21905 & 4.769432 \\
    \hline

    \multirow{3}{*}{$p^2(1/\upsilon,R)$} & \multirow{3}{*}{Traditional} & \multirow{3}{*}{3-1}  & Winner (W) & 0.922383 & 9.581754  \\ \cline{4-6}
                                             &                               &                      & Loser (L)  & -1.52129 & 14.81716 \\ \cline{4-6}
                                             &                               &                      & W - L      &  2.443672 & 6.01926 \\
    \hline

    \multirow{3}{*}{$p^3(1/\sigma,R)$} & \multirow{3}{*}{Traditional} & \multirow{3}{*}{3-1}  & Winner (W) & 0.65425 & 6.644175  \\ \cline{4-6} 
                                             &                               &                      & Loser (L)  & -0.61694 & 8.667792 \\ \cline{4-6}
                                             &                               &                      & W - L      &  1.27119 & 3.425697 \\
    \hline
    \end{tabular}
    \label{tab:yearly_returns}
\end{table}

\paragraph{}
Table~\ref{tab:yearly_returns} contains all long-horizon monthly mean returns and standard deviations of the best-performing portfolios from every mass and momentum combination. All the long horizon physical momentum portfolios have a higher mean return than the benchmark Nifty index, with $p^2(\upsilon,R)$ portfolio having the highest monthly mean return of 3.21\%. Additionally, the $p^2(\upsilon,R)$ portfolio has lower volatility (4.76\%) than the Nifty benchmark (6.10\%).

\begin{table}[]
    \caption{Risk measures of long horizon physical momentum portfolios in NSE 500.}
    \centering
    \begin{tabular}{|l|l|l|l|l|l|l|l|}
    
    \hline
    Portfolio & Strategy & J-K & Basket & Fin. Wealth & Sharpe Ratio & VaR$_{95\%}$ & MDD\\
    \hline
    
    $p^1(\upsilon,R)$ &	Traditional   & 8-1 & (W - L) & 2.016235 & 0.335661 & 4.357561 & -16.5675 \\ 
    \hline

    $p^1(1/\upsilon,R)$ & Traditional  & 6-2 & (W - L) & 2.171599 & 0.343956 & 4.397591 & -16.5916 \\ 
                
    \hline

    $p^2(\upsilon,R)$ &	Traditional   & 9-4 & (W - L) & 2.138261 & 0.319444 & 4.725888 & -16.2056 \\
    \hline

    $p^2(1/\upsilon,R)$ & Traditional & 6-2  & (W - L)  & 1.381071 & 0.109031 & 7.716793 & -31.127 \\
    \hline

    $p^3(1/\sigma,R)$ & Traditional & 12-4  & (W - L)   &  1.259924 & 0.135235 & 5.122885 & -21.3893 \\

    \hline
    \end{tabular}
    \label{tab:yearly_risks}
\end{table}

\paragraph{}
All of the best-performing portfolios are traditional establishing a continuation in long-horizon returns. $p^3(1/\sigma,R)$ momentum is the only long horizon portfolio whose monthly mean return failed to outperform the benchmark in terms of mean, however, $p^3(1/\sigma,R)$ had the least standard deviation as well. P1 and P2 portfolios had similar performance in terms of mean returns, however, P1 portfolios performed better than P2 portfolios in terms of standard deviation. Additionally, P1 and P2 portfolios' gains were primarily attributable to the loser group's significant continuation of returns, whilst their respective winner groups demonstrated weak continuation. The low performance of the P3 portfolio can be attributed to the weak continuation of physical momentum in its winner and loser groups. It can also be seen that the dollar-neutral portfolios exhibit lower volatilities when compared to the individual winner and loser groups that constitute them. This reduction in volatility can be attributed to the diversification effect introduced by the assets of the winner and loser groups.

\paragraph{}
Table~\ref{tab:yearly_risks} contains the risk measures of all the long-horizon dollar-neutral portfolios. $p^3(1/\sigma,R)$ and $p^2(1/\upsilon,R)$ are the only portfolios that failed to outperform Nifty in every risk measure. The momentum portfolios’ 95\% VaR values range from 4.35\% to 7.12\% and their maximum drawdowns range from -16.20\% to -31.12\%. For P1 portfolios maximum drawdowns reduced by nearly 43\% when compared to the benchmark Nifty. The failure of $p^3(1/\sigma,R)$ and $p^2(1/\upsilon,R)$ portfolios can be attributed to higher drawdowns and VaR measures.

\begin{figure}[ht]
\centering 
{\includegraphics{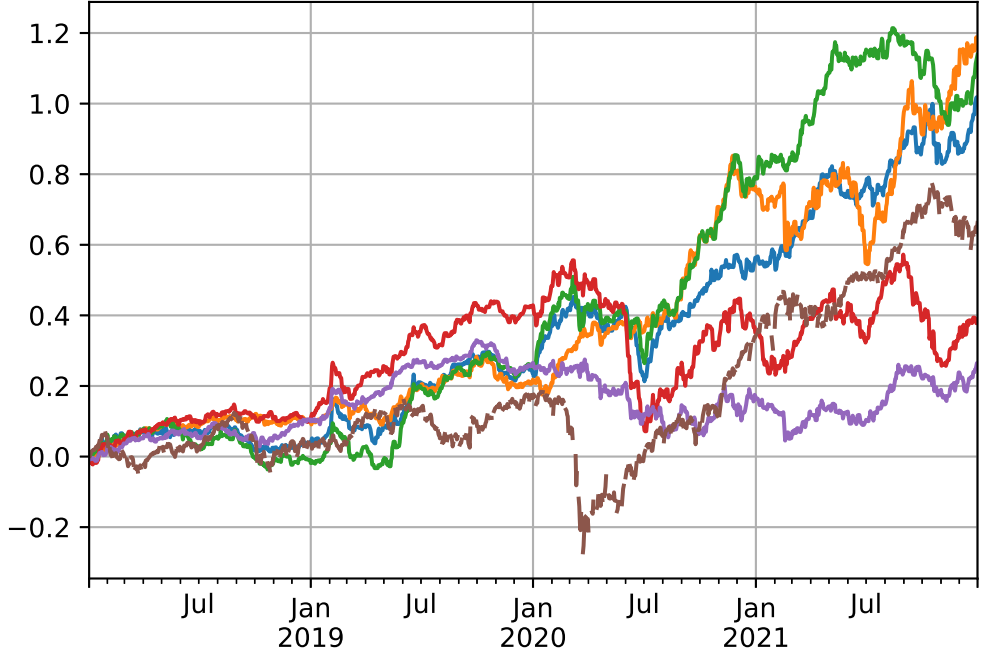}}
\caption{Cumulative returns for $p^1\left(\upsilon,R\right)$ (Blue), $p^1\left(1/\upsilon,R\right)$ (Yellow), $p^2\left(\upsilon,R\right)$ (Green), $p^2\left(1/\upsilon,R\right)$ (Red), $p^3\left(1/\sigma,R\right)$ (Purple) long horizon physical momentum portfolios vs Nifty 50 (Brown) in Indian Market.}

\label{fig:yearly}
\end{figure}

\paragraph{}
Figure~\ref{fig:yearly} shows the historical yearly portfolio values of all long-horizon momentum portfolios along with the Nifty benchmark. The P3 momentum portfolio was the worst-performing portfolio in terms of final wealth, returning 1.2 times the initial investment. Whereas $p^1(1/\upsilon,R)$ is the best-performing portfolio yielding a 2.17-fold return on the initial investment. From the figure above, it can be seen that prior to the Covid-19 market crash in March 2020, all momentum portfolios did outperform the benchmark Nifty, but the major outperformance of $p^1(\upsilon,R)$, $p^1(1/\upsilon,R)$, and $p^2(\upsilon,R)$ occurred during the post-Covid-19 bull run observed in the Indian market \cite{dhall2020covid}. $p^2(1/\upsilon,R)$ portfolio was the most affected by the Covid-19 market crash observing a drawdown of nearly 32\% during this period. Additionally, $p^2(1/\upsilon,R)$ and $p^3\left(1/\sigma,R\right)$ portfolios were the only portfolio that failed to leverage the post-covid-19 bull run and hence failed to outperform the benchmark Nifty.

\section{Conclusion}
In this paper, we apply the physical momentum strategy, whose quantitative and mathematical definitions are introduced in Choi (2014), to the Indian Market. We created momentum portfolios based on a ranked selection of winner and loser stocks from the NSE 500 index. Stocks comprising the winner and loser baskets are chosen based on their physical price momentum values. All the physical momentum portfolios outperform the benchmark Nifty 50 in terms of expected returns and risk profiles. The majority of these physical momentum portfolios are traditional, indicating that the winner basket outperforms the loser basket, implying a robust continuation of stock returns in the Indian market. High-frequency portfolios (daily) are the only exception where the loser baskets outperform. All of the physical momentum portfolios have sharpe-ratio values of less than one, indicating that they are not suitable for investment. Given the assets in our portfolios are equally weighted, the Sharpe ratio can be improved by optimizing the weights of each asset. The application of the Fama-French model to explain the performance of the physical momentum portfolios is beyond the scope of this research and will be addressed in future work. 

\bibliographystyle{unsrt}
\bibliography{references}

\end{document}